\documentstyle[preprint,aps,floats]{revtex}
\input epsf

\newcommand{\ket}[1]{|{#1}\rangle}
\newcommand{\braket}[2]{\langle #1 \ket{#2}}

\newcommand{\aga}{Al$_{0.5}$Ga$_{0.5}$As}
\newcommand{\gip}{Ga$_{0.5}$In$_{0.5}$P}
\newcommand{\aia}{Al$_{0.5}$In$_{0.5}$As}
\newcommand{\Eq}[1]{Eq.~(\ref{#1})}
\newcommand{\Fig}[1]{Fig.~\ref{#1}}
\newcommand{\mc}[2]{\multicolumn{#1}{#2}}

\begin{document}
\draft
\preprint{TP-451-6221, to be published in Appl.\ Phys.\ Lett.}
\title{Effects of atomic clustering on the optical properties of
III-V alloys}
\author{Kurt A. M\"ader and Alex Zunger}
\address{National Renewable Energy Laboratory, Golden, CO 80401}
\date{\today}
\maketitle
%
\begin{abstract}
Self-consistent electronic structure calculations together with a
structural model are used to study the effect of short-range atomic
order on the optical properties of otherwise random \aga, \gip, and
\aia\ alloys. We find that clustering can reduce the
direct band gap of these alloys by as much as 100~meV. Furthermore,
sufficiently strong clustering is predicted to transform \aga\ into a
direct gap material.
\end{abstract}
%
\nobreak

\narrowtext

Pseudobinary $A_{1-x}B_xC$ semiconductor alloys generally exhibit
deviations from perfect random arrangements of the $A$ and $B$ atoms
on their fcc sublattice. These deviations take the form of
long-range order (LRO), short-range order (SRO), or both.  LRO in
III-V alloys appears most frequently in the CuPt structure and is
accompanied by a reduction in the band gap relative to the disordered
phase\cite{Zunger93}. This gap reduction reflects zone-folding and
level-repulsion \cite{Wei89} and depends quadratically \cite{Laks92}
on the degree $\eta$ of LRO. In contrast, studies of the effects of
SRO on alloy band gaps are scarce.  The degree of SRO is generally
quantified by the Warren-Cowley
\cite{Cowley50} parameter
\begin{equation}
\alpha_j = 1 - \frac{P_B(j)}{x_B},
\label{def:alpha}
\end{equation}
where $P_B(j)$ is the probability to find a $B$ atom on the $j$-th
nearest-neighbor shell about $A$ as an origin. In the perfect random
alloy $P_B(j) = x_B$ and thus $\alpha_j = 0$ for all atomic
shells. Preferred association of {\em like atoms} (``clustering'') yields
$\alpha_j > 0$, while association of {\em unlike atoms} (``anticlustering'')
is manifested by $\alpha_j < 0$.
Direct measurements of SRO in tetrahedral semiconductor alloys were
carried out by diffuse x--ray scattering
\cite{Bruhl77,Yasuami92},
transmission electron microscopy
\cite{Gowers83}, and
scanning tunneling microscopy \cite{Sale93}. Indirect evidence for SRO
comes from nuclear
magnetic resonance \cite{Zax93}, resonant Raman scattering
\cite{Manor93},
infrared reflectivity \cite{Perk91}, deep impurity photoluminescence
\cite{Shira88}, and photoreflectance \cite{Dimoulas90}.
Many experiments
\cite{Bruhl77,Yasuami92,Gowers83,Sale93,Perk91,Dimoulas90} report
clustering-type SRO ($\alpha>0$), whereas anti-clustering seems to be
less frequently observed
\cite{Zunger93,Zax93}.
Despite these extensive studies, little is known about the effects of
SRO on band gaps
\cite{Fu89,Magri91}.  In this Letter we report results of
first-principles pseudopotential calculations on the band gaps of
perfectly random ($\alpha=0$) as well as clustered ($\alpha>0$) models
of \aga, \gip\ and \aia\ alloys. We find that local clustering
can (i) reduce the band gap of III-V alloys to a similar extent as LRO
does, (ii) transform the indirect-gap material \aga\ into a
direct-gap one, and (iii) localizes the band edge wavefunctions
preferentially on a particular type of cluster, which thus acts as a
``local quantum-well.''  We will discuss the chemical trends for
wavefunction localization in the sequence Al/Ga/In.

The standard approach to the electronic structure of alloys---the
virtual crystal approximation---does not distinguish $A$ from $B$
atoms irrespective of their chemical disparity.
The single-site coherent potential
approximation assumes that all $A$ atoms (and separately, all $B$
atoms) have the same charge transfer and bond sizes, irrespective of
their local environments.
A more realistic description would allow for
a {\em distribution} of $A$ and $B$ atoms reflecting the existence of
many distinct local atomic environments in the alloy.  The most direct
approach to this description would involve application of band theory
to fictitious solids with huge ($\gtrsim 1000$ atom) supercells
\cite{Li92,Hass90,Mader93} whose sites are occupied by $A$ and $B$ atoms
according to a prescribed degree (zero or finite) of SRO. This
``direct approach'' has been implemented for random 1000--2000 atom
models of
\aga\ using empirical tight-binding \cite{Hass90} and empirical
pseudopotentials in a plane-wave basis \cite{Mader93}.  These studies
showed that the results can be mimicked very well using much smaller
unit cells ($\sim$10 atoms) but with {\em specially selected site
occupations and cell geometries} (``special quasirandom structures'',
or SQS) \cite{Zunger90}.  We find that the error of the SQS energy
gaps relative to huge supercells is \cite{Mader93} $\sim$20 meV.  In
this work we use these SQSs in the context of self-consistent
non-local pseudopotential calculations within the local density
approximation (LDA) to obtain the electronic energy bands of III-V
alloys with and without SRO.  The SRO assumed is of the clustering
type in which the nearest-neighbor shell has $\alpha_1>0$ and all
subsequent shells are random.  We use 16-atom SQSs with the same
computational parameters as described in Ref.\
\onlinecite{Magri91}.
The atomic positions of \gip\ and \aia\ in the SQSs were relaxed using
a Keating-type valence force field \cite{Martin70}, while maintaining
cubic symmetry for the cell-external degrees of freedom.
We are interested in calculating alloy band gaps of a particular
zincblende representation, e.g., $\Gamma_{1c}$ or $X_{1c}$.  We
therefore average the eigenvalues $E_j(\bar{\bbox K})$ of a few SQS
states $\ket{j\bar{\bbox K}}$, weighted with the
spectral density of a zincblende-type state $\ket{n\bbox k}$ of a given
representation ($n$ and $j$ are band indices):
\begin{equation}
\label{eq:ave}
\langle E_n(\bbox k)\rangle_{\rm alloy} = \frac{1}{W} \sum_{j,\bar{\bbox K}}\;
|\braket{j\bar{\bbox K}}{n\bbox k}|^2 E_j(\bar{\bbox K}),
\end{equation}
where the sum runs over a few SQS states around a peak in the spectral
density of the state $\ket{n\bbox k}$, $\bar{\bbox K}$ must differ from
$\bbox k$ by a reciprocal lattice vector of the SQS, and $W$ is a
normalization constant.

Table \ref{tab:gaps} summerizes our calculated [\Eq{eq:ave}] alloy
band gaps
\cite{LDAcorr} with
and without SRO. We choose in this study a relatively large and
positive SRO parameter of $\alpha=\frac{1}{6}$ in order to emphasize
the effect of local clustering. (Note, however, that SRO parameters as
large as the present one have been reported in the literature
\cite{Bruhl77}).  The Table also gives the average band gaps of the
binaries $\bar{E_g} =
\frac{1}{2} [E_g(AC) + E_g(BC)]$ and the calculated gaps of CuPt ordered
alloys ($\eta=1$).
The optical bowing parameter $b$ is defined by $E_g(x) =
\bar{E_g}(x) - b x (1-x)$.
{}From the table we conclude the following:
(i) The direct band gaps of the
random alloys ($\alpha=\eta=0$) are reduced relative to $\bar E_g$
by 0.05, 0.10, and 0.13 eV for \aga, \gip, and
\aia, respectively. This gives
bowing parameters of 0.20, 0.40, and 0.52 eV, compared with the
experimental results of 0.37 (Ref.\ \onlinecite{Lee80}), 0.70 (Ref.\
\onlinecite{String72}), and 0.74 eV (Ref.\ \onlinecite{84W}),
respectively.
(ii) LRO of the CuPt type ($\eta=1$) reduces the  band
gaps relative to $\bar E_g$ by 0.41, 0.59, and 0.29 eV, respectively.
(iii) Local clustering with a SRO parameter $\alpha=\frac{1}{6}$ also reduces
significantly the direct band gaps relative to $\bar E_g$, i.e., by 0.14, 0.23,
and 0.22 eV for \aga, \gip, and \aia, respectively.
In
\aga\ the $\Gamma_{1c}$-like transition now lies {\em lower} in energy than
the $X_{1c}$-like indirect transition. The latter seems to be almost
unaffected by SRO.  Thus we predict an indirect to direct band gap
crossover in \aga\ with sufficient local clustering.  This is in
marked contrast to the findings of Fu et.\ al.\ who find {\em
negative} optical bowing upon clustering \cite{Fu89}. Magri et.\ al.\
obtain the same sign of band gap modification as we do, but a much
smaller value due to the smaller SRO parameter assumed there
\cite{Magri91}.

In order to understand the physical mechanism leading to the large gap
reduction reported in Table \ref{tab:gaps}, we analyze the
wavefunctions of the lowest conduction (CBM) and highest valence (VBM)
states in the absence and then in the presence of SRO.  We find that
as SRO sets in, the wavefunctions at either the lowest conduction band
edge (\aga) or the highest valence band edge (\aia) or both (\gip)
strongly localize on clusters of one type of cations, while the
corresponding states in the random alloys show no such tendency.  This
is illustrated by plotting contours of the wavefunctions squared in
crystallographic planes intersecting these clusters (see right hand
side of Fig.~\ref{fig:psi}).  The wavefunctions are clearly segregated
on a particular cation sublattice, indicated in the fourth column of
Table~\ref{tab:gaps}.  The {\em degree} of segregation (i.e.,
localization) can be quantified by projecting the wavefunctions on
atomic spheres belonging to a particular type of cation tetrahedron
$A_{n}B_{4-n}$. We define a cluster weight as
\begin{equation}
\label{eq:proj}
w_n(j\bar{\bbox K}) = \frac{1}{N_n} \sum_{\bbox r_i} |\braket{\chi_\Omega(\bbox
r_i)}{j\bar{\bbox K}}|^2,
\end{equation}
where $N_n$ is the number of clusters of type $A_{n}B_{4-n}$, the sum
runs over all positions of $A,B$, and $C$ atoms participating in such a
cluster, and $\chi_\Omega(\bbox r_i)$ is a characteristic function
which is zero outside an atomic sphere of volume $\Omega$ centered at
$\bbox r_i$, and $1/\Omega$ inside the sphere.
Histograms of the cluster weights $w_n$ for the conduction band
minimum of \aga, \gip, and the valence band maximum of
\aia\ are also shown in Figure~\ref{fig:psi}, demonstrating localization
on particular ``pure'' clusters.
(The CBM in \aia\ is weakly localized on In-rich clusters).

The results of \Fig{fig:psi} imply that the clusters act as
``impurity-like traps'' for electrons or holes, very much like
isoelectronic impurities which are known to bind carriers if the
difference between their local potential and that of the host atom
exceeds a critical value
\cite{Bald72}.  However,
whereas an {\em isolated} Ga impurity is not strong enough to bind an
electron in either a AlAs or InP host crystal
\cite{Hjal80}, the formation of clusters in the respective
alloys spatially extends the range of the perturbing potential, thus
enhancing its strength and leading to binding.  Similarly, an isolated
In impurity in GaP or AlAs will probably not bind a hole, but the
corresponding clusters do.  For sufficiently large cluster sizes we
can rephrase the preceding argument in terms of band theory: the
segregation of the band edge wavefunctions is dictated by the band
offsets of the binary components taken at the alloy volume $\bar V =
V(x=0.5)$.  For example, whereas unstrained InP has a much smaller
band gap than unstrained GaP, hydrostatic compression of the former
and expansion of the latter to their common alloy volume $\bar V$
reverses the order of the lowest conduction states, so now Ga
forms the conduction band edge.  (The valence band maximum in
compressed InP is only $\sim$0.25 eV higher \cite{Wei-private} than the one
in expanded GaP).  Thus, the
lowest conduction wavefunction in a clustered \gip\ alloy should
segregate on the Ga-rich clusters as found in the calculation
(Fig.~\ref{fig:psi}).
For electrons, the observed trend follows the order of atomic
$s$-orbital binding energies, which increase in the sequence
Ga$\to$In$\to$Al (LDA values are $-$9.16, $-$8.46, and $-$7.83 eV,
respectively).
The hole localization, however, is less directly correlated with an
atomic property of the cations alone, since the wavefunctions at the
VBM are delocalized on the {\em anion sublattice}.

In summary, we have shown that local cation clustering in common-anion
alloys can reduce the energy gap considerably with respect to the
ideal random alloys.  This is accompanied by localization
of the band edge wavefunctions on clusters with lower
potential energy.  It would be interesting to study experimentally
the band gaps of clustered III-V alloys.  It is conceivable that
intentional creation of SRO may be used as a tool for band gap
engineering.

This work was supported by the office of Energy Research, Materials
Science Division, U.S.\ Department of Energy, under grant No.\
DE-AC02-83CH10093.

%

%
%
%
\begin{figure}[t]
\hbox to \hsize{\epsfxsize=0.50\hsize\hfil\epsfbox{figure-1.eps}\hfil}
\nobreak\bigskip
\caption{The left hand side
histograms represent the weights (Eq.\ \protect\ref{eq:proj}) that CBM
or VBM wavefunctions have on the $A_nB_{4-n}$ clusters in the presence
of clustering SRO.  The atomic sphere volumes $\Omega$ used for the
projections are equal to the average atomic volume in each alloy.  In
the right hand side we show contour plots of the corresponding wave
function squared on the (110) plane. The states are clearly segregated
on one type of cluster.}
\label{fig:psi}
\end{figure}

%
\newcommand{\Gc}{\Gamma_{1c}}
\newcommand{\Xc}{X_{1c}}
\begin{table}
\caption{LDA corrected band gaps in eV (measured from the
top of the valence band) for three alloy systems at different
states of order: perfect randomness ($\alpha=\eta=0$), clustering-type
SRO ($\alpha=\frac{1}{6}$), and CuPt-type LRO ($\eta=1$).  Here,
$\bar{E_g}$ denotes the average gap of the binaries at their
equilibrium volumes.  Chemical symbols in parentheses denote the
sublattice on which the VBM and CBM are localized, respectively, and D
denotes that the state is delocalized.}
\label{tab:gaps}
\begin{tabular}{llllllll}
\mc{2}{l}{System} & \mc{1}{c}{$\bar E_g$} & \mc{1}{c}{random} &
	\mc{2}{c}{SRO} & \mc{2}{c}{LRO}\\
                  &  & & & \mc{2}{c}{$\alpha=\frac{1}{6}$} &
	\mc{2}{c}{$\eta=1$} \\
\hline
\aga    & $\Gc$ &  2.27  & 2.22 & 2.13 & (Ga/Ga) & 1.86 & (D/Ga) \\
        & $\Xc$ &  2.18  & 2.17 & 2.16 &         & 2.10\\
\gip    & $\Gc$ &  2.16  & 2.06 & 1.93 & (In/Ga) & 1.57 & (In/Ga) \\
\aia    & $\Gc$ &  1.78  & 1.65 & 1.56 & (In/D)  & 1.49 & (In/D) \\
\end{tabular}
\end{table}


\begin{references}
\bibitem{LDAcorr} The LDA energy eigenvalues are rigidly shifted to agree with
experiment for each binary constituent. For the alloys we use the average of
these shifts, which are 0.84, 0.61, and 0.77 eV, for \aga, \gip, and \aia,
respectively. Note that energy {\em differences} such as optical bowing
coefficients are not affected by this procedure.
\end{references}
\end{document}